\documentclass[aps,twocolumn,prl,amsmath,amssymb,showpacs,superscriptaddress]{revtex4-1}

\usepackage{graphicx}
\usepackage{color}

\makeatletter
\@ifundefined{textcolor}{}
{
 \definecolor{BLACK}{gray}{0}
 \definecolor{WHITE}{gray}{1}
 \definecolor{RED}{rgb}{1,0,0}
 \definecolor{GREEN}{rgb}{0,1,0}
 \definecolor{BLUE}{rgb}{0,0,1}
 \definecolor{CYAN}{cmyk}{1,0,0,0}
 \definecolor{MAGENTA}{cmyk}{0,1,0,0}
 \definecolor{YELLOW}{cmyk}{0,0,1,0}
}

\begin{document}

\title{Temperature induced Lifshitz transition in WTe$_2$}

\author{Yun Wu}
\affiliation{ Ames Laboratory, U.S. DOE and Department of Physics and Astronomy, Iowa State University, Ames, Iowa 50011, USA}

\author{Na Hyun Jo}
\affiliation{ Ames Laboratory, U.S. DOE and Department of Physics and Astronomy, Iowa State University, Ames, Iowa 50011, USA}

\author{Masayuki Ochi}
\affiliation{ RIKEN Center for Emergent Matter Science (CEMS), Wako, Saitama 351-0198, Japan}
\affiliation{ JST ERATO Isobe Degenerate $\pi$-Integration Project, Advanced Institute for Materials Research (AIMR), Tohoku University, Sendai, Miyagi 980-8577, Japan}

\author{Lunan Huang}
\affiliation{ Ames Laboratory, U.S. DOE and Department of Physics and Astronomy, Iowa State University, Ames, Iowa 50011, USA}

\author{Daixiang Mou}
\affiliation{ Ames Laboratory, U.S. DOE and Department of Physics and Astronomy, Iowa State University, Ames, Iowa 50011, USA}

\author{Sergey~L.~Bud'ko}
\affiliation{ Ames Laboratory, U.S. DOE and Department of Physics and Astronomy, Iowa State University, Ames, Iowa 50011, USA}

\author{P. C. Canfield}
\email[]{canfield@ameslab.gov}
\affiliation{ Ames Laboratory, U.S. DOE and Department of Physics and Astronomy, Iowa State University, Ames, Iowa 50011, USA}

\author{Nandini Trivedi}
\affiliation{Department of Physics, The Ohio State University, Columbus, OH 43210, USA}

\author{Ryotaro Arita}
\affiliation{ RIKEN Center for Emergent Matter Science (CEMS), Wako, Saitama 351-0198, Japan}
\affiliation{ JST ERATO Isobe Degenerate $\pi$-Integration Project, Advanced Institute for Materials Research (AIMR), Tohoku University, Sendai, Miyagi 980-8577, Japan}

\author{Adam Kaminski}
\email[]{kaminski@ameslab.gov}
\affiliation{ Ames Laboratory, U.S. DOE and Department of Physics and Astronomy, Iowa State University, Ames, Iowa 50011, USA}

\date{\today}

\begin{abstract}
We use ultra-high resolution, tunable, VUV laser-based, angle-resolved photoemission spectroscopy (ARPES) and temperature and field dependent resistivity and thermoelectric power (TEP) measurements to study the electronic properties of WTe$_2$, a compound that manifests exceptionally large, temperature dependent magnetoresistance. The temperature dependence of the TEP shows a change of slope at T=175~K and the Kohler rule breaks down above 70-140~K range. The Fermi surface consists of two electron pockets and two pairs of hole pockets along the X-$\Gamma$-X direction. Upon increase of temperature from 40K, the hole pockets gradually sink below the chemical potential. Like BaFe$_2$As$_2$, WTe$_2$ has clear and substantial changes in its Fermi surface driven by modest changes in temperature. In WTe$_2$, this leads to a rare example of temperature induced Lifshitz transition, associated with the complete disappearance of the hole pockets. These dramatic changes of the electronic structure naturally explain unusual features of the transport data.
\end{abstract}
\pacs{}
\maketitle

The discovery of giant magnetoresistance (GMR) in Fe/Cr superlattice\cite{Baibich88PRL,Binasch89PRB}, 
opened a new era of applications in magnetic field sensors, read heads in high density hard disks,
random access memories, and galvanic isolators\cite{Daughton99JMMM}. 
In the quest of achieving large MR in crystalline materials, a class of manganese oxides colossal 
magnetoresistance (CMR) materials was discovered that exhibits a large change in resistance with 
applied magnetic fields but only at low temperatures\cite{Urushibara95PRB,Moritomo96Nat,Ramirez97Nat}. 
Recently, extremely large MR has been observed in PtSn$_4$\cite{Mun12PRB}, Cd$_3$As$_2$\cite{Liang15NatMat}, NbSb$_2$\cite{Wang14SciRep}, and WTe$_2$\cite{Ali14Nat}. In both PtSn$_4$ and WTe$_2$, the MR shows no sign of saturation and reaches an order of at least $10^5$\% at low temperature. Different mechanisms have been proposed to explain the large MR in these materials, such as high electron/hole mobility and carrier density in PtSn$_4$\cite{Mun12PRB}, protection mechanism from electron backscattering in Cd$_3$As$_2$\cite{Liang15NatMat}, electron-hole carrier compensation in WTe$_2$\cite{Ali14Nat,Pletikosic14PRL,Alekseev15PRL}. However, the exact origin 
of large MR in these materials remains an open question.

The temperature dependent electrical resistivity, Hall coefficient and thermoelectric power of tungsten-ditelluride have been known for several decades \cite{Kabashima66JPSJ} and a three-carrier semi-metal band 
model \cite{Brixner62JINC,Kabashima66JPSJ} was proposed to explain the electrical resistivity. 
Later on, density-functional based augmented spherical wave (ASW) electronic structure calculations 
and relatively low resolution ARPES \cite{Augustin00PRB} were used to study the electronic properties of WTe$_2$ and further supported the semimetallic nature of this material. However due to the low resolution of these early experiments, no details about the Fermi surface or band dispersion were obtained. Recent ARPES \cite{Pletikosic14PRL} and quantum oscillation \cite{KamranPRL15} results  revealed presence of small electron and hole pockets of roughly similar size. These findings were consistent with carrier compensation mechanism as the primary source of the MR effect \cite{Ali14Nat,Alekseev15PRL}. Furthermore, this study also reported a change of the size of the Fermi pockets between 20K and 100K.  More recently, Jiang {\it et al.} \cite{Jiang15arXiv}  claimed observation of strong spin-orbital coupling effect and proposed that the backscattering protection mechanism also may play a role in the large non-saturating MR of WTe$_2$. 

\begin{figure*}[htbp]
	\includegraphics[width=4in]{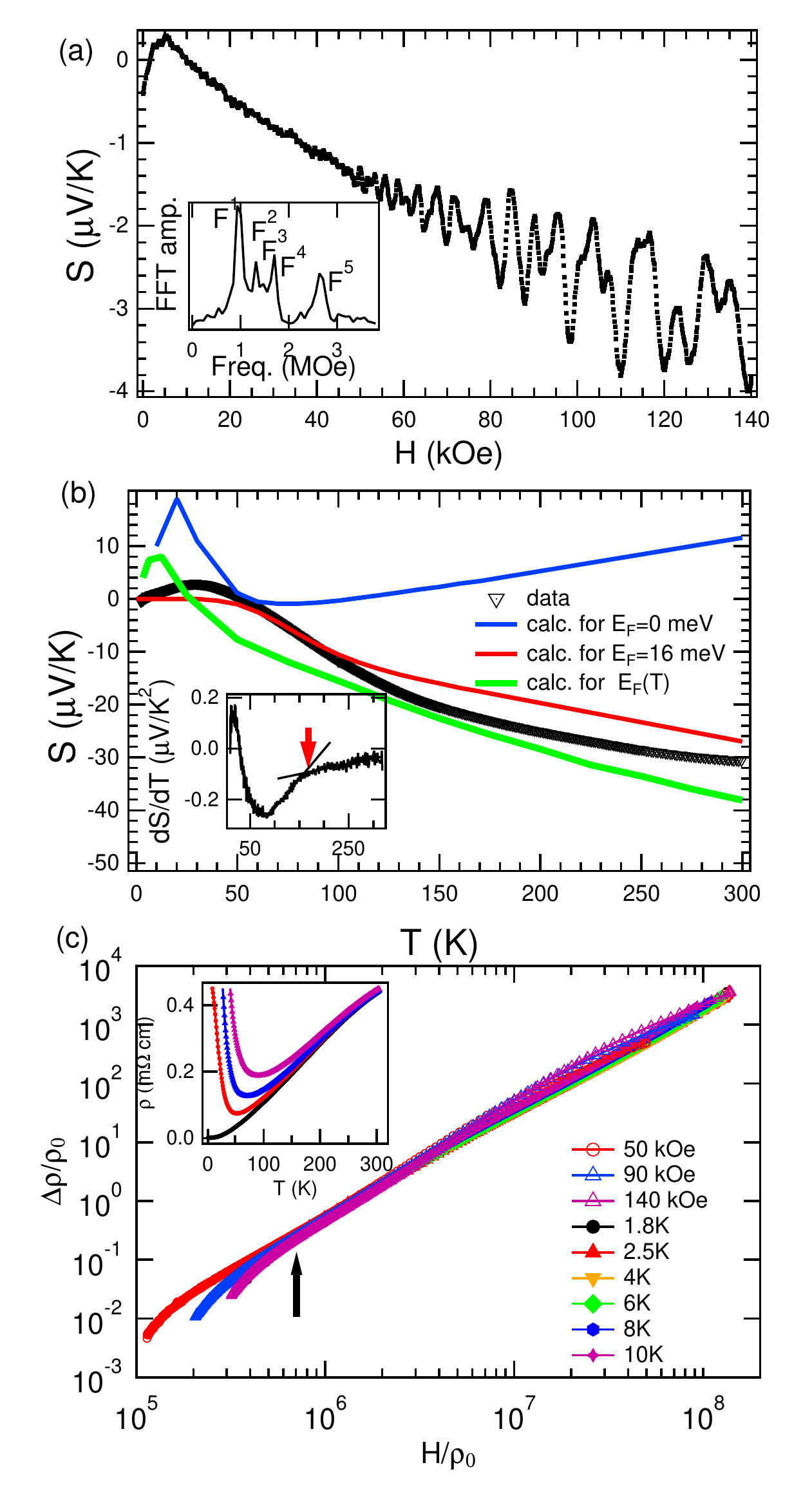}%
	\caption{(color online) (a) Magnetic field dependence of TEP measured at T=2.2K with very clear quantum oscillations. Inset shows FFT of data after subtraction of a smooth background. 
	(b) Temperature dependence of TEP. Inset shows first derivative with arrow marking the change of slope due to Lifshitz transition. Blue line and red lines are calculated x component of TEP assuming E$_F$=0, and E$_F$=20 meV. Green line shows calculated x component of TEP assuming that E$_F$ changes from 0 meV at T=0 to 45 meV at T=300~K.
	(c) Generalized Kohler plot. Arrow marks the point below which the Kohler rule is violated ($T<60$ K). Inset shows temperature dependence of the resistance measured for magnetic field of 0 kOe, 50 kOe, 90 kOe and 140 kOe.
		\label{fig:Transport}}
\end{figure*}

In this letter, we present the results from temperature dependent transport and ultra-high resolution, tunable VUV laser based ARPES\cite{Jiang14RSI} measurements. Our data show that there are two electron pockets and four hole pockets along the X-$\Gamma$-X direction 
and a fully occupied light ``hole" band at the center of the Brillouin Zone (BZ) located just below the Fermi energy. Systematic temperature dependence measurements revealed a Lifshitz transition, i. e. change of the Fermi surface topology, close to 160~K above which both pairs of the hole pockets vanish. This transition is likely associated with change of slope observed in the derivative of the temperature dependent thermoelectric power (TEP).

Whereas most of the previous measurements have been carried out on WTe$_2$ crystals grown via chemical vapor transport using halogens as transport agents\cite{Brown1966, Ali14Nat}, we have grown WTe$_2$ single crystals from a Te-rich binary melt.  High purity, elemental W and Te were placed in alumina crucibles in W$_1$Te$_{99}$ and W$_2$Te$_{98}$ ratios.  The crucibles were sealed in amorphous silica tubes and the ampoules were heated to 1000$^{\circ}$C over 5 hours, held at 1000$^{\circ}$C for 10 hours, and then slowly cooled to 460$^{\circ}$C over 100 hours and finally decanted using a centrifuge\cite{Canfield92PMPB}. The resulting crystals were blade or ribbon like in morphology with typical dimensions of 3 $\times$ 0.5 $\times$ 0.01 mm with the crystallographic $c$ axis being perpendicular to the larger crystal surface; the crystals are readily cleaved along this crystal surface. 

Temperature and field dependent transport measurements were performed in Quantum Design Physical 
Property Measurement system for 1.8 K $\leq$ T $\leq$ 350 K and $\vert$H$\vert$ $\leq$ 140 kOe.  The thermoelectric power (TEP) measurements were performed by a dc, alternating temperature gradient technique\cite{Eundeok10MST}. Temperature and field dependent resistivity measurements made on our solution grown samples (see Figs SI xx-zz) demonstrate exceedingly high values of the residual resistivity ratio (RRR) in excess of 900 and MR values at 1.8 K and 90 kOe up to 6x10$^5~$\%.  

Samples were cleaved \textit{in situ} at 40 K in ultrahigh vacuum (UHV). The data were acquired using 
a tunable VUV laser ARPES system, consisting of a Scienta R8000 electron analyzer, picosecond Ti:Sapphire oscillator and fourth harmonic generator\cite{Jiang14RSI}. Data were collected with a tunable photon energies ranging from 5.3 eV to 6.7 eV. Momentum and energy resolution were set at $\sim$ 0.005 \AA$^{-1}$ and 2 meV. The  size of the photon beam on the sample was $\sim$30 $\mu$m. 

For first-principles band structure calculations, we used the Perdew-Burke-Ernzerhof parametrization of the generalized gradient approximation~\cite{PBE} and the full-potential (linearized) augmented plane-wave plus local orbitals (FP-(L)APW+lo) method including the spin-orbit coupling as implemented in the \textsc{wien2k} code~\cite{wien2k}. Experimental crystal structure taken from Ref.~[\onlinecite{Expt_latt}] was used. The muffin-tin radii for W and Te atoms, $r_{\rm W}$ and $r_{\rm Te}$, were set to 2.4 and 2.38 a.u., respectively. The maximum modulus for the reciprocal lattice vectors $K_{\rm max}$ was chosen so that $r_{\rm Te} K_{\rm max}$ = 9.00. TEP was calculated using a 52$\times$29$\times$13 $k$-point mesh with the \textsc{BoltzTrap} code~\cite{BoltzTrap}.

The field dependence of the TEP at 2.2 K is shown in Fig.\ref{fig:Transport}(a).  As was the case for PtSn$_4$\cite{Mun12PRB}, very clear quantum oscillations can be seen in the TEP of WTe$_2$.  FFT analysis (Fig.\ref{fig:Transport}(a) inset) gives F$^1$ = 0.93 MOe, F$^2$ = 1.31 MOe, F$^3$ = 1.47 MOe and F$^4$ = 1.70 MOe in excellent agreement with the values found from Shubnikov-de Hass data (Fig. S2 in Supplemental Material).  The peak labeled F$^5$ is thought to be a result magnetic breakdown of F$^1$ and F$^4$ low field orbits\cite{KamranPRL15}.
The temperature dependence of the TEP is plotted in Fig.\ref{fig:Transport}(b). On warming TEP first increases with temperature reaching local maximum at $\sim$30K, then decreases, crossing zero near 50K and its absolute value continues to increase rapidly up to $\sim$160K. For higher temperatures the TEP changes at lower rate reaching value of {-~30~$\mu$V/K} at room temperature. The change of the rate at $\sim$160K is illustrated in the inset of \ref{fig:Transport}(b), where first derivative of TEP displays clear kink marked by an arrow. 
The temperature and field dependence of the extraordinarily large MR of WTe$_2$ are shown in \ref{fig:Transport}(c).  The generalized Kohlers plot shows that there is fairly good scaling of the data with an exponent of $\sim$1.98 at lower temperatures.  As temperature is increased Kohlers law scaling breaks down.  Given that temperature is an implicit variable, the temperature values associated with the vertical arrow in \ref{fig:Transport} c range from 70~K to 140~K.

\begin{figure*}[tb]
	\includegraphics[width=5in]{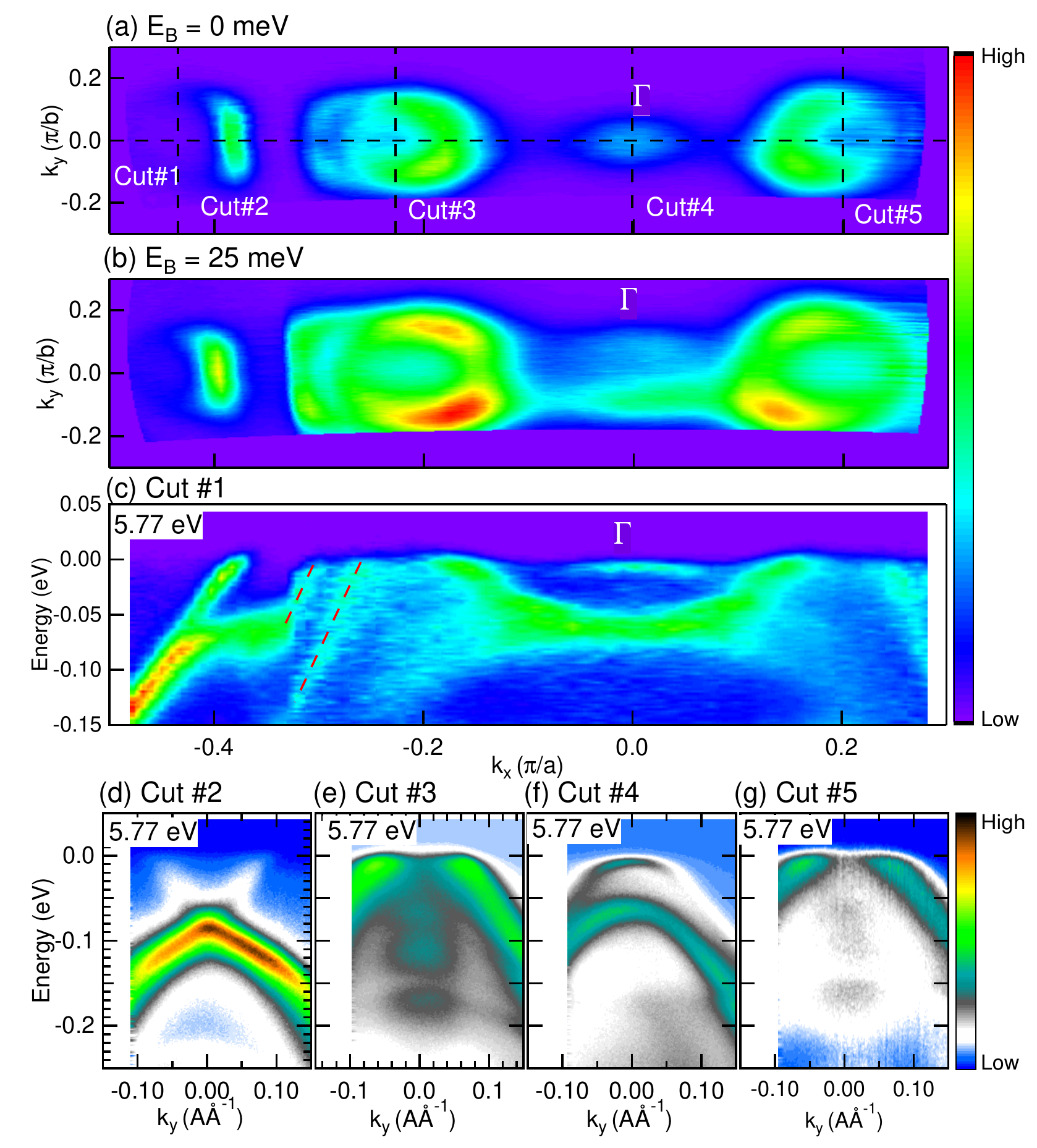}%
	\caption{(color online) Fermi surface plot and band dispersion. 
	a) Fermi surface plot - ARPES intensity integrated within 10 meV about the chemical potential.
	b) Constant energy plot at binding energy of 25 meV. 
	c) Band dispersion along cut \#1. Dashed lines mark the two left branches of the two hole bands.
	d) Band dispersion at the center of the electron pocket along cut \#2.
	e) Band dispersion at the center of the left hole pocket along cut \#3.
	f) Band dispersion at the center of the zone along cut \#4.
	g) Band dispersion at the center of the right hole pocket along cut \#3.
	\label{fig:ARPES_FS}}
\end{figure*}

\begin{figure*}[tb]
	\includegraphics[width=7in]{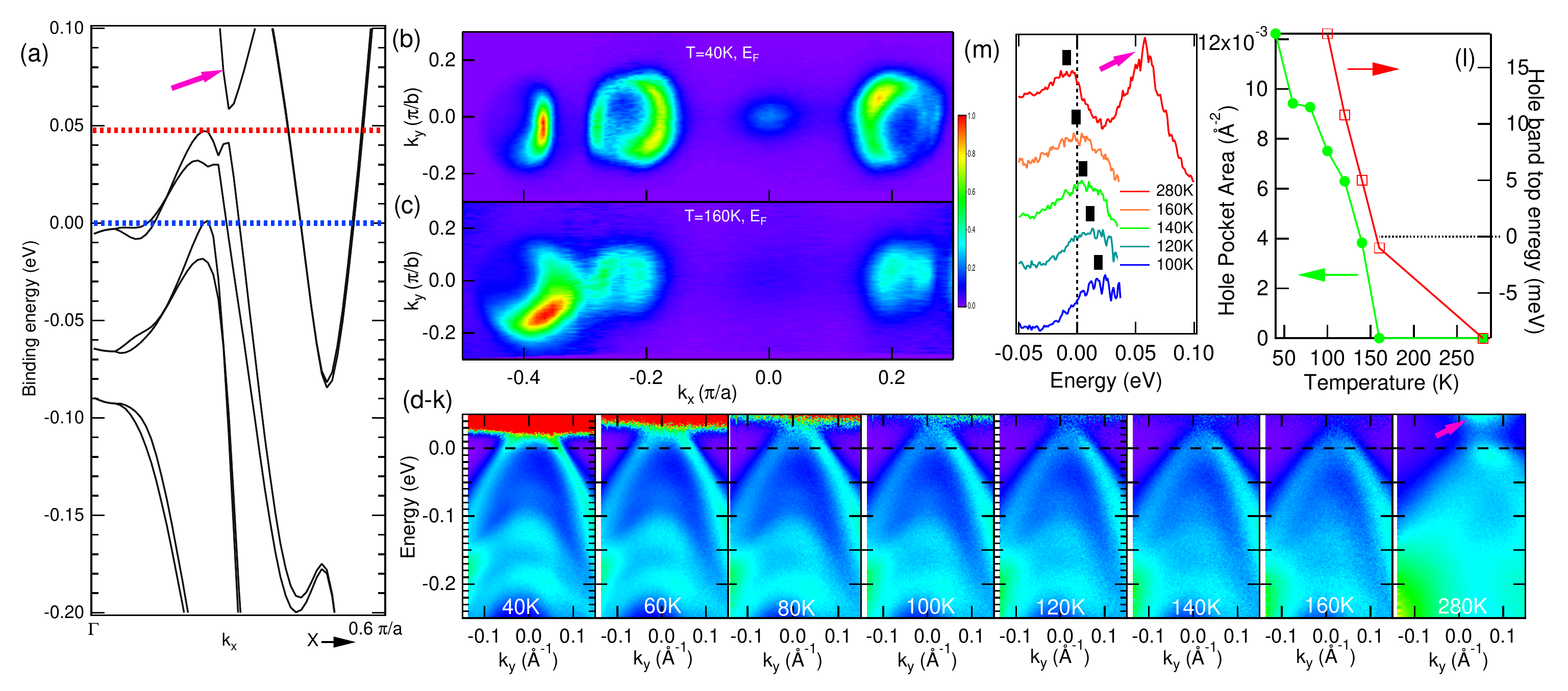}%
	\caption{(color online) (a) Calculated band structure along $\Gamma$-X symmetry direction. Blue and red dashed lines mark the values of chemical potential used for calculating TEP.
	b) Fermi surface plot - ARPES intensity integrated within 10 meV about the chemical potential measured at T=40K.
	c) same as (a) but for T=160K
	d-k) temperature dependence of the band dispersion at the hole pocket (i. e. along cut \#1 in Fig. 1a) divided by the Fermi function. 
	l) EDCs divided by Fermi function at the center of the hole pocket for several temperatures. Black line marks the energy of the peak.
	m) Temperature dependence of the area of the hole pocket and energy of the top of the hole band. 
Purple arrows in (a, l and k) point to a band located above the hole band that is identified in calculation, band dispersion and Fermi function divided EDCs. 
	\label{fig:ARPES_T}}
\end{figure*}

We now proceed to elucidate the electronic origin of the strange transport behavior i. e. change of the slope of the TEP and violation of the Kohler's rule. The details of the Fermi surface and band dispersion along key directions in the Brillouin zone are shown in Fig.\ref{fig:ARPES_FS}.  In Fig.\ref{fig:ARPES_FS}(a) we plot the ARPES intensity integrated within 10 meV about the chemical potential, where high intensity contours mark the location of the 
Fermi surface sheets. The Fermi surface consists of a pair of hole pockets and an electron pocket
on each side of the $\Gamma$ point. The electron pocket on the right side is not visible as it is 
outside of the scan range. The intensity at binding energy of 25 meV is shown in Fig.\ref{fig:ARPES_FS}(b). 
The size of the contours for the electron pockets decreases and size of the contour of the hole pockets 
increases demonstrating the correct assignment of the character of the carriers. 
The band dispersion along the main symmetry axis is plotted in Fig.\ref{fig:ARPES_FS}(c). Due to 
matrix elements, only the right branch of the electron band is clearly visible. This is followed by two 
crossings of the left sides of the hole bands (marked by dashed lines), and then a single crossing of 
the right branch of both hole bands. At the center of the BZ, the top of hole band is located just below E$_F$, thus there is no hole pocket at the center of the zone. Detailed band dispersion along vertical cuts are shown in Fig.\ref{fig:ARPES_FS}(d-g). Fig.\ref{fig:ARPES_FS}(d) shows that the bottom of the electron band joins with the top of a lower band and appears just like the structure of a Dirac state\cite{Geim07NatMat} approximately 70 meV below E$_F$. This is different from calculations shown in Fig.~ \ref{fig:ARPES_T}a, where bottom of the electron pocket is separated from the band below by a 200 meV gap. Cut \#3 and Cut \#5 show the dispersion of the hole pockets. At the cut location, the bands that belong to this pair of hole pockets seem already merged within our resolution. At the $\Gamma$ point (Cut \#4), two bands are visible with the top one located just below E$_F$.

We now proceed to describe one of more intriguing electronic property of this material that leads to unusual transport properties. 
Fig.\ref{fig:ARPES_T} a shows the calculated band structure along $\Gamma$-X direction. The band calculation predicts pair of hole pockets and an electron pocket between $\Gamma$ and X, in agreement with ARPES data presented above and previous calculations\cite{Ali14Nat}.
In Figs.\ref{fig:ARPES_T}b, c we show the Fermi surface map measured at 40K and 160K. The hole pockets shrink from two circles to a spot of intensity
and electron pocket expands with increasing temperature. This is consistent with previous comparison of low and high temperature FS size \cite{Pletikosic14PRL}. 
We detail this behavior by plotting  ARPES intensity divided by the Fermi function along the vertical cut at the center of the hole pocket  in Fig.\ref{fig:ARPES_T}(d-k). A clearly visible hole band moves down in energy and by 160~K its top 
 touches the chemical potential and at temperature of 280~K, the top of this band has sunk 
below the chemical potential. %
To quantify this effect we plotted the EDCs divided by the Fermi function at the center of the hole pocket for several temperatures in Fig.\ref{fig:ARPES_T}(l). 
We extract the energy position of the peak by fitting a gaussian and plot such extracted energy location of the top of the hole band in Fig.\ref{fig:ARPES_T}(m). The top of the hole band moves down in energy upon increasing temperature from 18 meV at 120~K to -7 meV at 280~K. We also extracted the area of the hole pocket by measuring the separation between MDC peaks as a function of temperature and plot this quantity in Fig.\ref{fig:ARPES_T}(m). The energy position of the top of the hole band moves below the chemical potential and thus the area of the hole pocket must vanish above $\sim$160~K.  
These data provide an archetypical example of temperature induced Lifshitz transition since they demonstrate a change of the Fermi surface topology upon heating. Chemical substitution induced Lifshitz transitions are quite common and were previously observed in Ba(Fe$_{1-x}$Co$_{x}$)$_2$As$_2$ at Co concentrations of 3.8\% 11\% and 20\% \cite{Liu10NatPhys, Liu11PRB}. On the other hand, a temperature induced Lifshitz transition in the absence of a structure or magnetic phase transition is extremely rare. (Interestingly enough  in Ba(Fe$_{1-x}$Ru$_{x}$)$_2$As$_2$, where the Fermi surface consists of hole and electron pockets similar, significant shifts of the chemical potential and resulting changes in size of both pockets were observed\cite{Dhaka13PRL}.) The temperature dependent TEP,  in particular the change of slope of $\partial S/\partial T$ at $\sim 160$ K (Fig. 1, inset) is consistent with the suggested temperature induced Lifshitz transition, as TEP is expected to be very sensitive to the changes in the Fermi surface topology \cite{varlamov}.

Whereas the Lifshitz transition itself can be inferred from both the TEP and ARPES data as outlined above, the dramatic, temperature dependent change in relative size of the electron and hole pockets should manifest itself in other ways over wider temperature ranges. 
 Blue and red curves in \ref{fig:Transport} b show calculated x-component of TEP using two fixed values of E$_F$ of 0 meV and 20 meV respectively. Neither of these two curves matches the experimental data even on qualitative level. The E$_F$=0 meV scenario (blue line) has positive peak at low temperature, but also positive values in high temperature limit unlike seen in the experimental TEP data. The E$_F$=20 meV curve has the correct high temperature limit, but lacks positive peak at low temperature. Clearly, calculation with fixed chemical potential does not reflect experimental data even on a qualitative level. The temperature induced shift of the chemical potential discussed above naturally solves this problem. In conventional metals, the chemical potential does not change with temperature appreciably at room temperature when the k$_B$T$\ll$E$_F$. In WTe$_2$ and other semimetals, where the top of the hole band and bottom of there electron band is in close proximity (few tens of meV) of E$_F$ one can expect significant changes of the chemical potential with temperature (also see calculated density of states presented in Supplemental Material). To demonstrate this we calculated the shift of the chemical potential with temperature by imposing requirement that the total number of electrons in all bands is conserved (see Supplemental Material). This leads to a change of the chemical potential of 14 meV between T=0~K and T=300~K. We repeated calculation of TEP using such obtained temperature dependent value of the chemical potential scaled so that $E_F=$45 meV at T=300 K to account for possible renormalization effects. The high and low temperature values of the chemical potential used in calculation of TEP are illustrated by dashed lines in Fig.\ref{fig:ARPES_T} a. 
The result of the calculation is plotted as green line in Fig. 1 b. Although this is an approximation and does not take into account the thermal expansion and other effects such as phonon drag \cite{Barnard} it does reproduce key features of the TEP data on qualitative level including positive peak at low temperature and correct trend in high temperature limit.
The MR data presented in Fig.~\ref{fig:Transport} c also indicates that a simple Kohlers rule model breaks down as the sample is heated.  Such a breakdown is likely associated with a changing ratio of electron and hole carriers implied by the data in \ref{fig:ARPES_T} m and caused by the temperature induced shift of the chemical potential.

In summary, we used ultra-high resolution tunable laser ARPES  to investigate the electronic structure of 
WTe$_2$. 
Fermi surface consists of two electron pockets and two pairs of hole pockets. This is consistent with the band calculation reported earlier\cite{Ali14Nat}. 
Upon increasing temperature, both electron and hole bands move to higher binding energy, or in 
other words the chemical potential is increasing. Eventually this leads to vanishing of 
the hole pocket for T$>$160K and thus a Lifshitz transition occurs, above which the Fermi surface 
consists of only two electron pockets. Around this temperature we observe discontinuity of the 
slope of the derivative of TEP, confirming that the Lifshitz transition affects the transport and thermodynamics 
of this material. This effect is also naturally explains features of temperature dependence of the TEP and breakdown of Kohler rule. We speculate that the temperature induced shift of the chemical potential and 
resulting change of the relative size of electron and hole pockets causes the large magnetoresistance 
effect to vanish at elevated temperatures. Reported here change of the chemical potential upon 
increase of the temperature is not expected for simple materials, but was observed previously 
in iron arsenic high temperature superconductors\cite{Dhaka13PRL}, where both electron and hole pockets are located in close proximity of E$_F$. 

We would like to thank Mohit Randeria for very useful discussions. Research was supported by the US Department of Energy, Office of Basic Energy Sciences, Division of Materials Sciences and Engineering. Ames Laboratory is operated for the US Department of Energy by the Iowa State University under Contract No. DE-AC02-07CH11358. Na Hyun Jo is supported by the Gordon and Betty Moore Foundation EPiQS Initiative (Grant No. GBMF4411).

\bibliography{WTe2}

\newpage

\begin{widetext}
\newpage

{\bf \noindent \fontsize{18}{24} \selectfont Supplementary material for "Temperature induced Lifshitz transition in WTe$_2$"}

\section{Transport measurements}

Resistance of the samples as a function of temperature and magnetic field (Fig. S\ref{Fig. S1.}) was measured in a standard, linear four probe configuration using the ACT option of a Quantum Design Physical Property Measurement System (PPMS) instrument. Clear Shubnikov - de Haas oscillations are seen in the magnetoresistance at low temperatures. For analysis, the oscillatory part of magnetoresistance (after subtraction of a monotonic part approximated by a second order polynomial) is plotted as a function of $1/H$ (Fig. S\ref{Fig. S2.}) and fast Fourier transform algorithm is used.

Hall resistivity, $\rho_H$, was measured using a four probe method, in positive  ($\rho_{+H}$) and negative  ($\rho_{-H}$) magnetic field and was calculated as $\rho_H = (\rho_{+H} - \rho_{-H})/2$ to separate odd in field $\rho_H$ from possible contamination of the even in field magnetoresistance. The results are presented in Fig. S\ref{Fig. S3.}.

\begin{figure*}[h]
		\centering
			\includegraphics[width=5in]{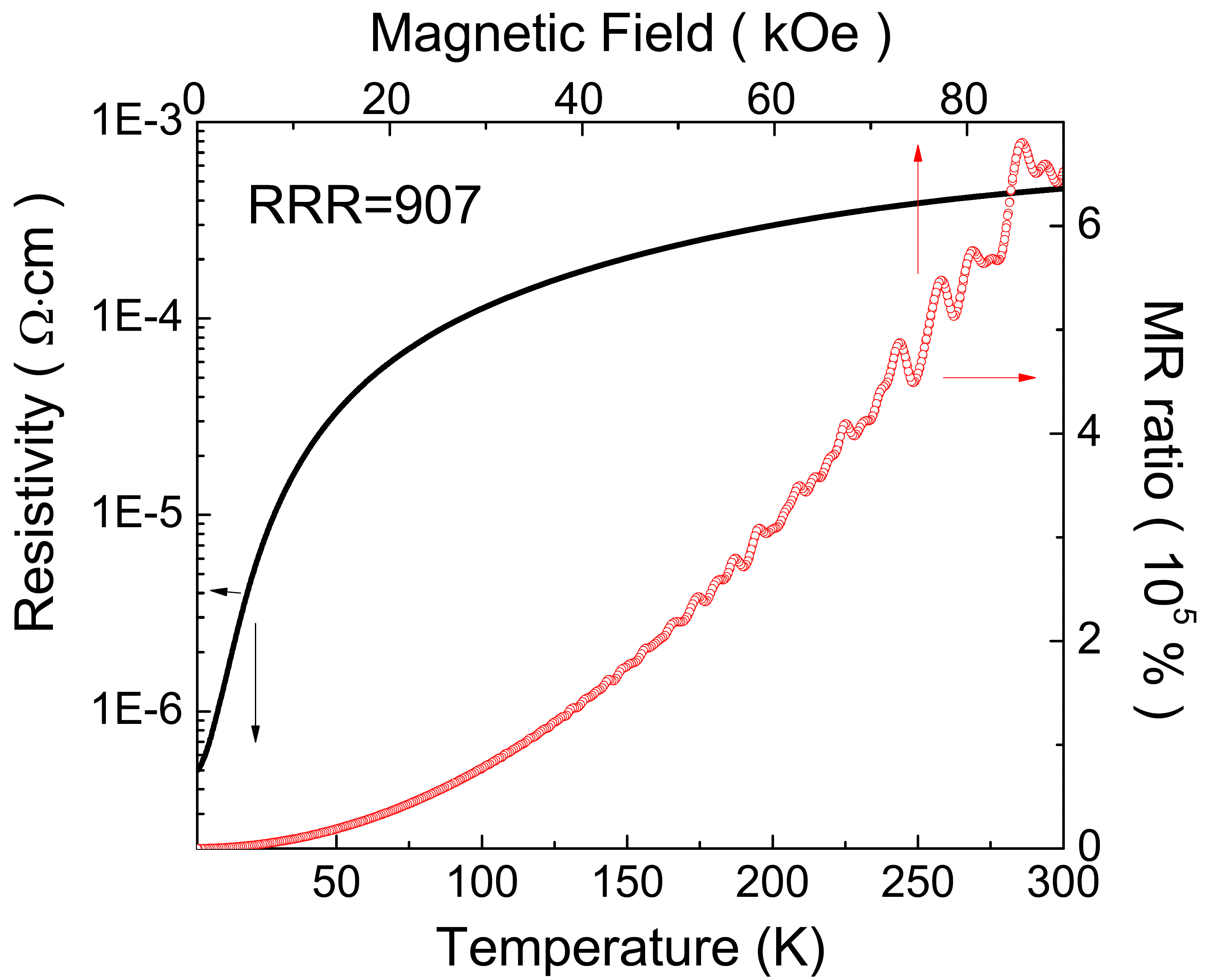}
		\caption{(color online) Resistivity data with residual resistivity ratio of$\sim 907$ and magnetoresistance of $\sim  6.5 \times 10^5 \%$ at 1.8 K and 90 kOe (H$\parallel$c).}
		\label{Fig. S1.}
	\end{figure*}

\begin{figure*}[h]
		\centering
			\includegraphics[width=5in]{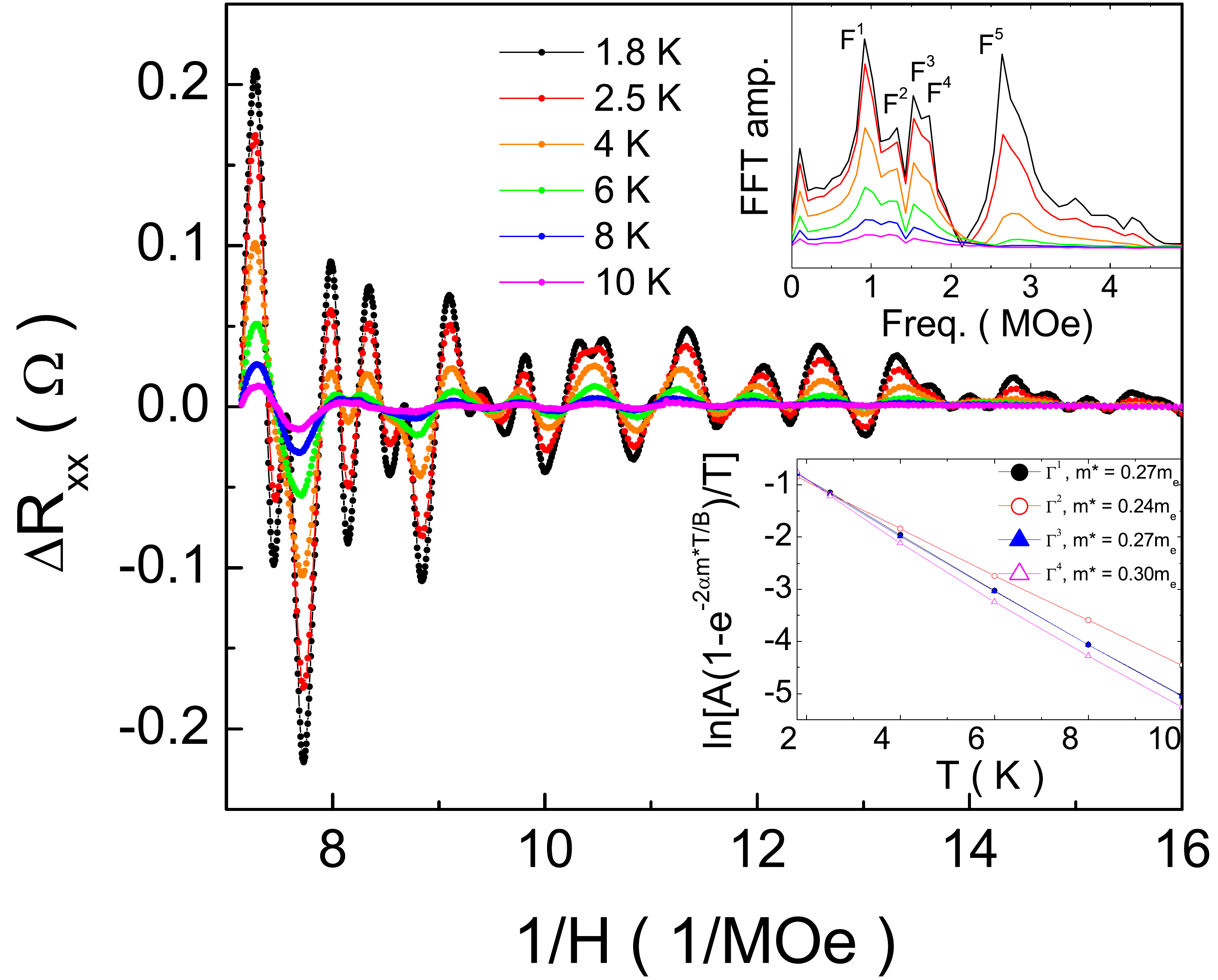}
		\caption{(color online) Shubnikov-de Haas oscillations for $H\parallel c$ at selected temperatures. Upper inset shows fast Fourier transform (FFT) spectra. It represents five different frequencies: $F^{1}=0.92~MOe$, $F^{2}=1.32~MOe$, $F^{3}=1.52~MOe$, $F^{4}=1.70~MOe$, $F^{5}=1.73~MOe$. The frequency, $F^{5}$, seems to be due to the magnetic breakdown of $F^{1}$ and $F^{4}$ orbits. The slope of the lower inset demonstrates effective mass of each frequencies based on Lifshitz-Kosevich theory, $m^{*}_{F^{1}}=0.27m_{e}$, $m^{*}_{F^{2}}=0.24m_{e}$, $m^{*}_{F^{3}}=0.27m_{e}$, $m^{*}_{F^{4}}=0.30m_{e}$.}
		\label{Fig. S2.}
	\end{figure*}
	
	\begin{figure*}[h]
		\centering
			\includegraphics[width=5in]{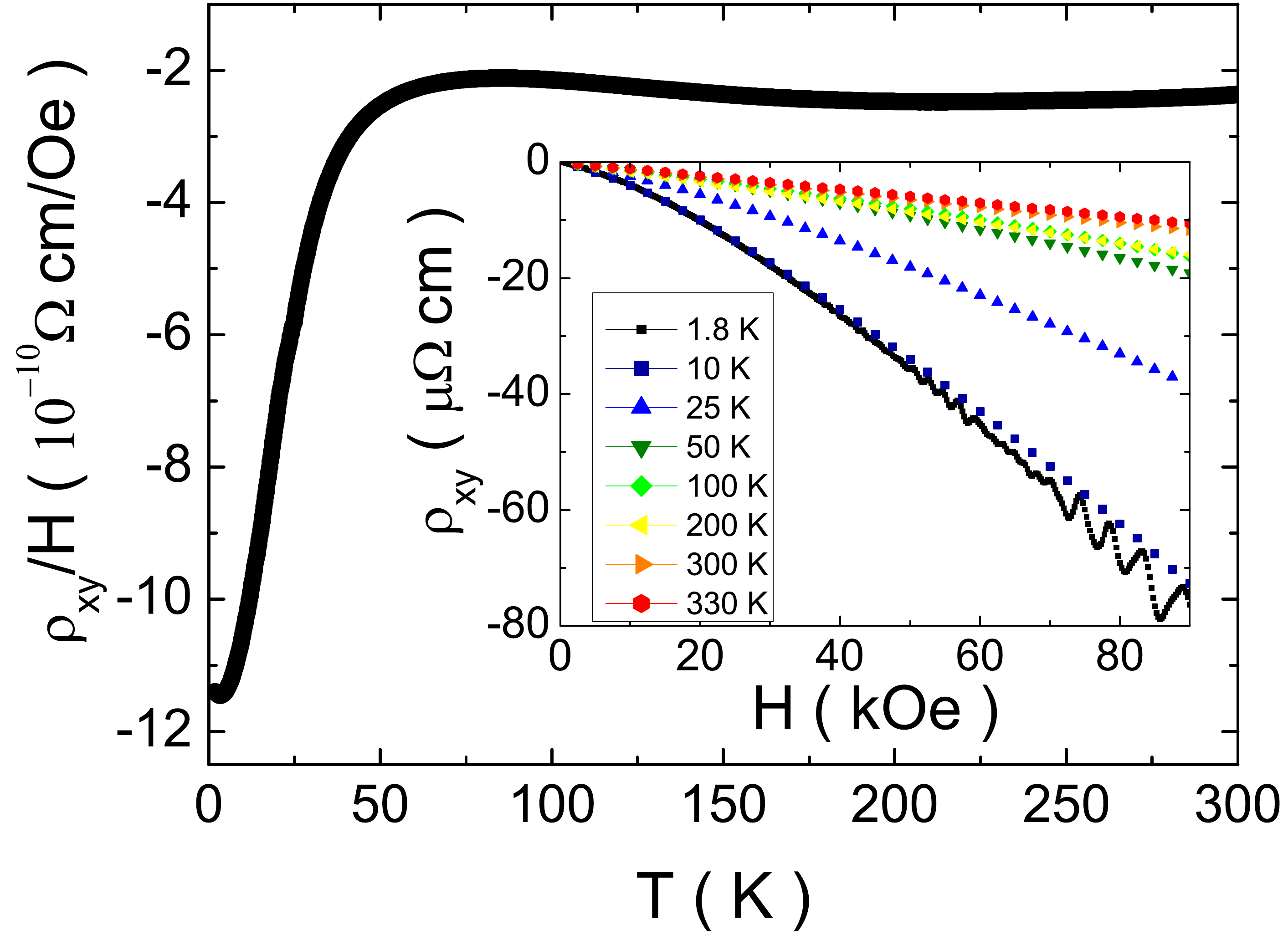}
		\caption{(color online) Hall coefficient, $R_{H}=\rho_{H}/H$, at H=90 kOe. Inset shows Hall resistivity of WTe$_{2}$ at T=1.8, 10, 25, 50, 100, 200, 300 and 330 K. Quantum oscillations in Hall resistivity are clearly seen at 1.8K}
		\label{Fig. S3.}
	\end{figure*}

\section{Theory Calculations}

The technical details of the band structure calculation are provided in main text. Fig. S4a shows calculated band structure along main symmetry directions. In panel Fig. S4b we plot the calculated density of states (DOS). The negative slope of the calculated DOS (i. e. DOS decreasing when increasing the energy) implies that the chemical potential will increase with increasing temperature and DOS in this temperature range is dominated by the hole pockets. 

The changes of the value of the chemical potential  with temperature are shown in Fig. S5 and were calculated by fixing the total number of electrons in all bands while applying a Fermi distribution for set temperature. 

\begin{figure*}[]
	\includegraphics[width=5in, angle =90]{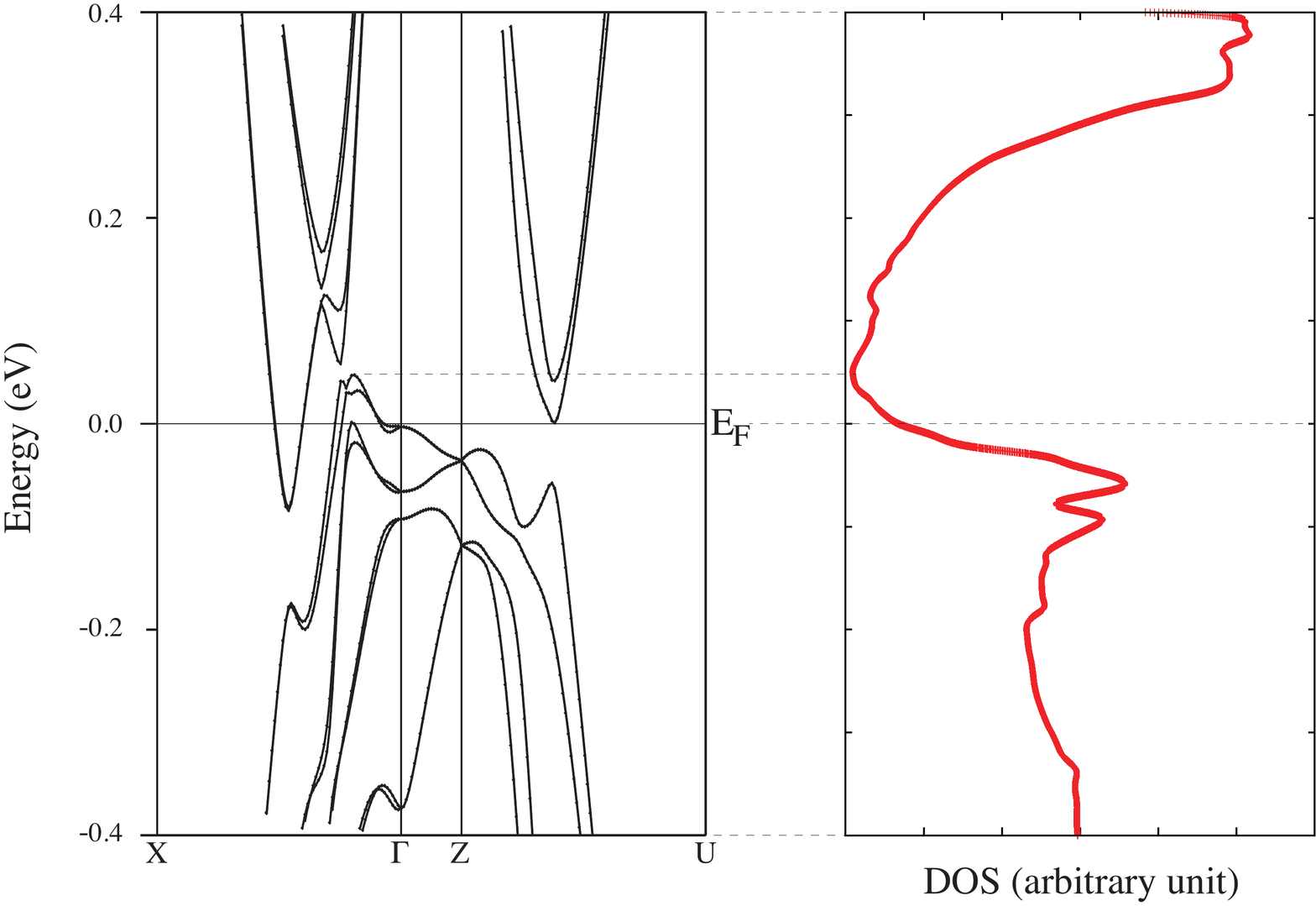}
	\caption{(color online) Calculated band structure (left) and corresponding density of states (right). The upper dashed line marks energy of the top of the hole band that corresponds to minimum in DOS, the lower dashed line marks $E_{F}$.}
	\label{fig:Band}
\end{figure*}

\begin{figure*}[]
	\includegraphics[width=5in]{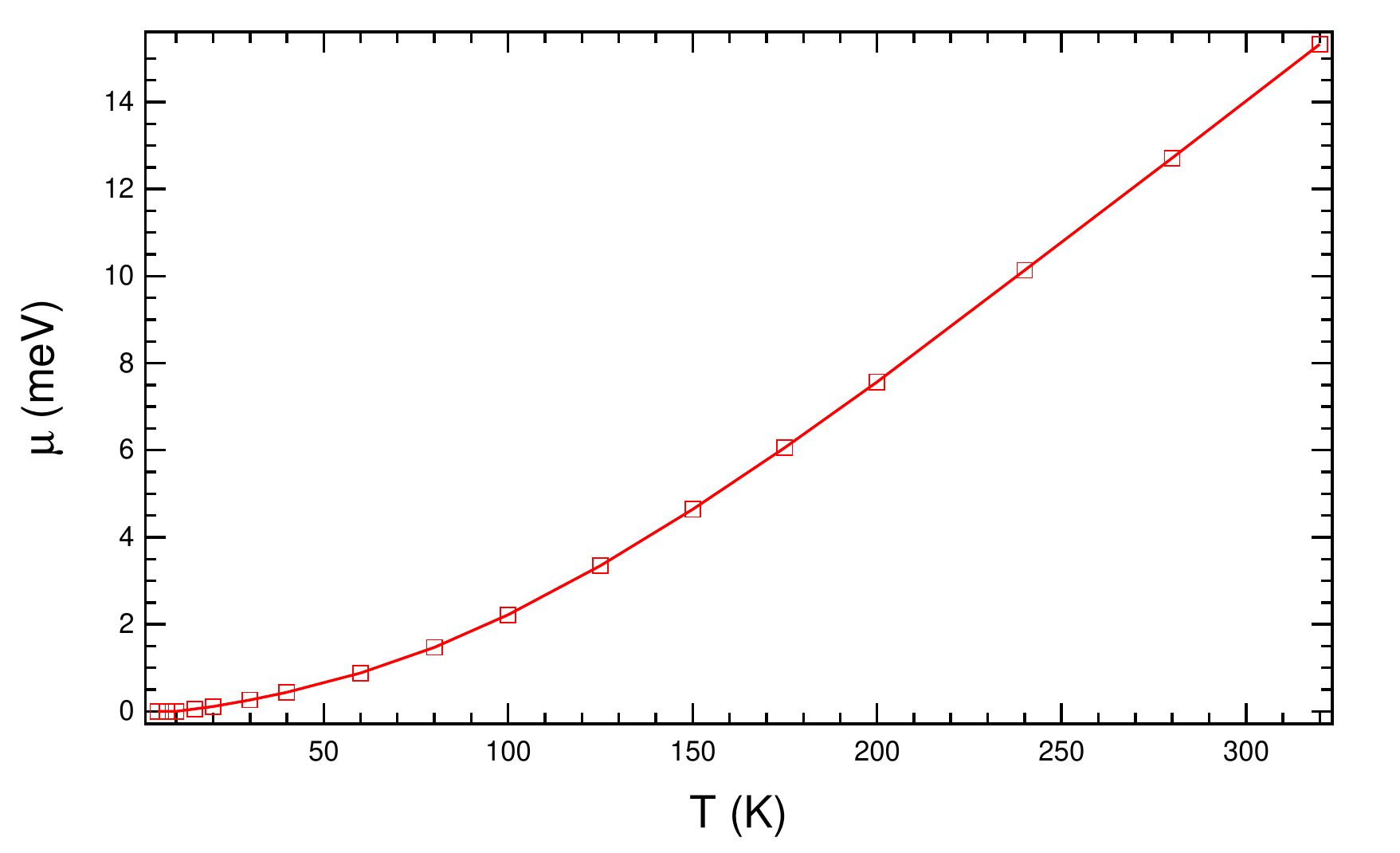}
	\caption{(color online) Calculated changes of the value of the chemical potential with temperature.}
	\label{fig:Band}
\end{figure*}
\end{widetext}

\end{document}